\documentstyle[aps,prl,epsf,graphics,multicol,times]{revtex}
\begin{document}
\draft
\title{Evidence of Two Distinct Dynamic Critical Exponents
  in Connection with Vortex Physics 
  }
\author {Petter Minnhagen and Beom Jun Kim}
\address {Department of Theoretical Physics,
Ume{\aa} University,
901 87 Ume{\aa}, Sweden}
\author{Hans Weber}
\address{Department of Physics,
  Lule{\aa} University of Technology,
  971 87 Lule{\aa}, Sweden}
\preprint{\today}
\maketitle
\thispagestyle{empty}
\begin{abstract}
The dynamic critical exponent $z$ is determined from
numerical simulations for the three-dimensional (3D) lattice Coulomb gas (LCG) and the 3D $XY$
models with relaxational dynamics. It is suggested that the dynamics
is characterized by two distinct dynamic critical indices
$z_0$ and $z$ related to the divergence of the relaxation time $\tau$ by
$\tau\propto \xi^{z_0}$ and $\tau\propto k^{-z}$, where
$\xi$ is the correlation length and $k$ the wavevector. The values
determined are $z_0\approx 1.5$ and $z\approx 1$ for the 3D LCG
and $z_0\approx 1.5$ and $z\approx 2$ for the 3D $XY$ model.
It is argued that the nonlinear $IV$ exponent relates to
$z_0$, whereas the usual Hohenberg-Halperin classification 
relates to $z$. Possible implications for the
interpretation of experiments are pointed out.
Comparisons with other existing results are discussed.
\end{abstract}


\begin{multicols}{2}
A neutral superfluid like $^4$He and
a superconductor 
in the limit of large London penetration depth can be characterized by a
complex order parameter.
The $XY$ model can be viewed as a discretized version
of this type of systems in terms of the phase of the order parameter~\cite{minnhagen_rev}.
An interesting feature of this class of models is the
presence of thermally generated topological defects which in three
dimensions (3D) take the form of vortex loops.
The superconducting
phase transition from the vortex loop point of view separates
a low-temperature phase consisting of closed vortex loops of
finite extent from a high-temperature phase where the loops
can disintegrate~\cite{williams1,williams2,williams3}.
The {\em static} thermodynamic properties of the $XY$
model allow a dual representation in terms of the vortex 
degrees of freedom. In particular the Villain transformation of the
$XY$ model gives rise to the dual vortex loop model termed the
lattice Coulomb gas (LCG)\cite{villain}. A crucial point in the
following is that this duality does not include the {\em dynamic}
properties which might be quite different.
However, all these models belong to the same
universality class for the static phase transition which means that
the static critical indices are the same.

The universality of the dynamic behavior is weaker and requires in
addition to the static universality also
that certain global features of the
dynamics are the same. In the dynamic universality
classes defined by Hohenberg and Halperin~\onlinecite{hohenberg} these
additional global features are expressed as local conservation laws.
According to this scheme the dynamics of a 3D
superfluid belongs to model F characterized by the dynamic critical exponent $z=1.5$~\cite{notez}.
A model with purely relaxational dynamics on the other hand belongs to
model A with $z\approx 2$~\cite{hohenberg}.
In case of a superconductor both model F and model A have been proposed
as the appropriate dynamical class~\cite{hohenberg,ffh}. 
Since relaxational dynamics is related to model A, it  
might appear surprising that the critical dynamic index $z$ for
the 3D LCG for periodic boundary conditions (PBC) and with purely relaxational dynamics in
Ref.~\onlinecite{weber} (verified in Ref.\onlinecite{wallin}) was
found to be $z\approx 1.5$ instead of $z\approx 2$.
Similarly the 3D $XY$ model with fluctuating twist boundary condition
(FTBC) and relaxational dynamics was found to have $z\approx
1.5$~\cite{melwyn}.
On the other hand, the 3D $XY$ model with PBC
and relaxational dynamics has $z\approx 2$ consistent with
model A.
This implies that the choice of boundary condition affects the value
of $z$~\cite{melwyn}. 
In the present Letter we show that the same type of sensitivity
applies to the 3D LCG, giving rise to two distinct exponents $z\approx
1.5$ and $z\approx 1$.
It is proposed that these results for the LCG model and the $XY$ model
reflect the existence of
two distinct critical indices corresponding to $\tau\propto \xi^{z_0}$
and $\tau\propto k^{-z}$, where $\tau$, $\xi$, and $k$ denotes
relaxation time, coherence length and wavevector, respectively.

The 3D LCG on a cubic lattice is defined by the Hamiltonian\cite{villain,jose,carneiro} 
\begin{equation}\label{eq:HLCG}
H=\frac{1}{2}\sum_{\alpha=1}^3\sum_{i,j}q_{\alpha i}W({\bf r}_i-{\bf r}_j)q_{\alpha j} ,
\end{equation}
where ${\bf q}_i\equiv ({q_{\alpha i}})$ 
represent the vortex line segment variables for each lattice site,
  one for each of the three lattice directions ${\bf e}_\alpha$,
  $q_{\alpha i}\in [0, 1, -1]$
  corresponding to no vortex segment, respectively to a segment with
  vorticity $1$ or
  $-1$
  between neighboring lattice sites and subject to the constraint that
  the sum
  of the $q_{\alpha i}$'s for the six directed links reaching each lattice site is
  zero.
  We use PBC and $W(r)$
  is the lattice version of the Coulomb interaction\cite{carneiro} 
\begin{equation}
W({\bf r})=\frac{1}{L^3}\sum_{{\bf k}}\frac{4\pi^2 e^{i{\bf k}\cdot{\bf r}}}{{\bf \kappa}^2({\bf k})} ,
\end{equation}
where $L$ is the total length of the lattice, the lattice constant is 1, and 
$\kappa_\alpha=2\sin (k_\alpha/2)$ where $\alpha=$1, 2, and 3. The relaxational dynamics is implemented by a
Metropolis Monte Carlo update where each complete update of the
lattice is
associated with one time unit (for details see Ref.\onlinecite{weber}).
The voltage across the sample ${\bf V}$ is proportional to the
expansion rate of the vortex loops 
${\bf V}\propto \frac{d}{dt} \sum_{{\bf r}} {\bf r}\times {\bf q}_{\bf r}$
        where ${\bf r}$ denotes the site positions on the lattice.
        The resistance $R$ can then be obtained from the voltage
        fluctuations
        through the Nyquist formula $R\propto \int_{-\infty}^{+\infty}dt\langle V(t)V(0)\rangle$.
        The scaling connection, $R\propto 1/\tau$, leads to the size
        scaling at $T_c$,
        $R\propto L^{-z_0}$, and the $\xi$-scaling of the
        resistivity,
        $\rho\propto \xi^{1-z_0}$ in the critical region~\cite{ffh}. The 3D LCG has a phase
        transition
        at $T_c\approx 3.003$ with $\nu\approx 0.67$~\cite{olsson1,hasenbusch}.
        Finite-size
        scaling at $T_c$ gives $z_0\approx 1.5$~\cite{weber,wallin}.
        In Fig.1 we demonstrate that the size converged $\xi$-scaling
        slightly above $T_c$, using $\xi \propto (T-T_c)^{-\nu}$, also
 gives $z_0\approx 1.5$. This
shows that $z_0\approx 1.5$ is not a finite-size effect caused by
the boundary, but is a bulk property characterizing the dynamics.

The LCG model defined with periodic boundary conditions (PBC) corresponds to the $XY$ model defined with the fluctuating twist boundary condition (FTBC)~\cite{olsson}. The
Hamiltonian for the latter model is given by 
\begin{equation}\label{eq:XYFTBC}
H(\theta_{\bf r},{\bf \Delta})= -\sum_{{\bf r}{\bf \alpha}}
\cos(\theta_{\bf r}-\theta_{{\bf r}+{\bf \alpha}}-{\bf e}_ {\alpha}\cdot{\bf \Delta}),
\end{equation}
where the $\alpha$ summation is over the three nearest
neighbors of ${\bf r}$ in the three lattice base directions.
The relaxational dynamics is, as for the LCG model, implemented
by a Metropolis Monte Carlo update where each complete update
of the lattice is associated with one time unit (for details see
Ref.\onlinecite{bjk}).
We use PBC for $\theta_{\bf r}$ which means that $\Delta_\alpha$
is
the average twist of the angle $\theta({\bf r})$ across the lattice in the
$\alpha$ direction. The fluctuations in
the
twist variable $\Delta_\alpha(t)$ are directly related to the voltage $V_\alpha$
across the sample by
$V_\alpha=-L\frac{d}{dt}\Delta_\alpha$~\cite{beom}.
The correspondence between LCG with PBC and the $XY$ model with FTBC
basically hinges on the fact that the models defined with these
boundary conditions allow for voltage fluctuations across the system.  
In contrast the usual PBC applied to the 3D $XY$ model is equivalent
to ${\bf \Delta}\equiv 0$ and the voltage fluctuations across the system is then no
longer properly described by the model~\cite{melwyn}.
The 3D $XY$ model with PBC, within the Villain approximation, corresponds
 to the 3D LCG model with PBC described by the {\em modified} Hamiltonian~\cite{vallat}
 [compare Eq.(\ref{eq:HLCG})]
 \begin{equation}\label{eq:HLCGM}
H=\sum_{\alpha=1}^3 \left[\frac{1}{2}\sum_{i,j}q_{\alpha i}W({\bf r}_i-{\bf r}_j)q_{\alpha j} + U(2\pi M_\alpha/L^2)\right] ,
\end{equation}
  where $U(\phi)=\sum_{m=-\infty}^\infty \exp [-(\phi+2 m)^2/2T]$ is the
 Villain function
 and $M_\alpha$ is given by ${\bf
 M}=\frac{1}{2}\sum_{{\bf r}}{\bf r}\times {\bf q}_{\bf r}$.
Since this Hamiltonian in Fourier space has the structure
$H\propto \sum_{{\bf k}\neq 0, \alpha} {\bf q}_\alpha({\bf k})W({\bf k})
{\bf q}_\alpha({-\bf k}) + \delta_{{\bf k} 0} U$
 the difference with Eq.(\ref{eq:HLCG}) is that in the modified
 Hamiltonian (\ref{eq:HLCGM}) the $k=0$ mode is suppressed.
 We focus on the scaling of the vorticity correlation function
\begin{equation}\label{eq:G}
G(k,t)=\frac{1}{{\bf k}^2L^3}\langle q_{\alpha {\bf k}}(t) q_{\alpha
  -{\bf k}}(0)\rangle  ,
\end{equation}
where ${\bf k}$ is perpendicular to the $\alpha$-direction. 
In particular $G(k=0,t)$ has the scaling form $G(k=0, t)\propto F(t\xi^{-z})/\xi$~\cite{melwyn}.
Figure 2 demonstrates that this scaling is well borne out with the
value $z\approx 1$. The data are from the critical region just above
$T_c$ where the data are size converged and $\xi\propto
(T-T_c)^{-\nu}$, which again
emphasize, that this is
a bulk property and not a property that vanishes for $L=\infty$. Thus, just as for the
3D $XY$ model in Ref.\onlinecite{melwyn}, we obtain two distinct values of
$z\approx 1.5$ and $z\approx 1$, by allowing and suppressing the $k=0$
mode. We again stress that the scaling and the exponents are
obtained from data which are independent of the system size.

As a resolution of this dichotomy we suggest that these models
are characterized by two distinct indices corresponding to
$\tau\propto \xi^{z_0}$ and $\tau\propto k^{-z}$.
Assuming such a resolution would mean that the vorticity correlation
function $G(k,t)$ in Eq.(\ref{eq:G}) in general scales as
$G(k,t)\propto F(tk^z, t\xi^{-z_0}, k\xi)/\xi$ which precisely at
$T_c$ for $k_{\rm min}=2\pi/L$ reduces to  
\begin{equation}\label{eq:kscale}
LG(k_{\rm min},t)=F(tk_{\rm min}^z,tL^{-z_0})  .
\end{equation}
Choosing $t=xk_{\rm min}^{-z}$ gives $F(x, xL^{z-z_0})$ which goes to
$F(x,\infty)$ for large $L$ when $z>z_0$ and $F(x,0)$ for
$z<z_0$. This means that in the scaling limit of large $t$ ($t\gg
\tau_0$ where $\tau_0$ is a microscopic characteristic time) we will
approach a scaling $\tilde{F}(x)$ with $x=tk_{\rm min}^z$ for large (small) $x$
when $z>z_0$ ($z<z_0$). Vice versa we will approach a scaling limit
$\hat{F}(x)$ with $x=tL^{-z_0}$ for large (small) $x$
when $z_0>z$ ($z_0<z$).

We first test the possibility of two distinct
indices for the 3D $XY$
model with FTBC [see Eq.(\ref{eq:XYFTBC})].
The resistance $R$ for the
3D $XY$ with FTBC is readily calculated from $\Delta_\alpha(t)$
(see \onlinecite{beom} for details). $G(k,t)$
is obtained by replacing
$q_{\alpha {\bf r}}$ with $\tilde{q}_{\alpha}({\bf r})\equiv \sin (\theta_{\bf r}-\theta_{{\bf r}+{\bf e}_\alpha})$
in Eq.(\ref{eq:G}) and multiplying by $k^2$~\cite{minnhagen1}:
\begin{equation}\label{eq:GXY}
G(k,t)=\frac{1}{L^3}\langle \tilde{q}_{\alpha {\bf k}}(t) \tilde{q}_{\alpha -{\bf k}}(0)\rangle  .
\end{equation}
Figure 3(a) and (b) demonstrate, by using the scaling form $R\propto L^{-z_0}f(L(T-T_c)^{-\nu})$  and
 $LG(k,t)=\tilde{F}(tk_{\rm min}^z)$,
 that for this model $z_0\approx 1.5$ whereas $z\approx 2$, confirming
 that $z_0\neq z$. For $z_0<z$ it should in principle also be possible to observe a
 crossover to $\tilde{F}(tL^{-z_0})$ for small enough arguments (but
 still with $t\gg\tau_0$), however this limit was not reached in the
 simulations\cite{notescale}.
 It is interesting to note that the exponent $z\approx 2$
 associated with $\tau\propto k_{\rm min}^z$ for the 3D $XY$ model
 with FTBC, within error bars, has the same value as the exponent $z_0\approx 2$ found for the
 3D $XY$ model with PBC~\cite{melwyn}.
 This suggests that the difference between FTBC and PBC arises
 from the difference in the treatment of the voltage fluctuations
 across
 the system, or equivalently the $k=0$ fluctuations:
 The $k=0$ fluctuations for FTBC are associated with $z_0\approx 1.5$.
 Changing to PBC suppresses these fluctuations and the $k=0$ fluctuations
 for
 PBC would then corresponds to the $k_{\rm min}=2\pi/L$-fluctuations for
 FTBC.
 This argument implies that $z_0$ for PBC and $z$ for FTBC should be
 identical, in accordance with our numerical data.
 
 Applying the same reasoning to the 3D LCG with PBC would mean that
 $z_0\approx 1.5$ and $z\approx 1$. Since in this case $z<z_0$, this
 would mean that $LG(k_{\rm min},t)$ should scale like $\tilde{F}(x)$
 with $x=k_{\rm min}^zt$ for small enough $x$ [compare
 Eq.(\ref{eq:kscale})]. Figure 4(a) shows that this prediction is
 consistent with the data and that the scaling curve approaches the
 correct asymptotic form $\tilde{F}(x)\propto 1/x^{1/z}$\cite{nytt} in the
 limit of small $x$. Figure 4(b) shows that the
 data for the largest converged $t$ values are instead consistent with
 the scaling $\tilde{F}(x=tL^{-z_0})$,
 which is also in accord with the scaling
 form given by Eq.(\ref{eq:kscale}). From this we conclude that the
 apparent non-uniqueness of the critical exponent $z$ for the 3D $XY$ model
 and the LCG with relaxational dynamics is consistent with the
 existence of two distinct critical indices $z_0$ and $z$ for the case when the
 models are defined so as to allow for voltage fluctuations across
 the system.
 
We have here proposed the existence of two indices
for relaxational dynamics in
3D. On the other hand the corresponding model with 
the resistively shunted Josephson junction dynamics (RSJ) has
$z_0=z$~\cite{melwyn,nytt}.
This model
is closely related to the $XY$ model and differs in that the dynamics is
subject to local current conservation, i.e., it has a local
conservation law which is not fulfilled by relaxational dynamics. 
 In 2D the whole low temperature phase below the Kosterlitz Thouless
transition is quasicritical and for the 2D $XY$ model (with FTBC)
$z_0(T)=2\pi\Upsilon(T)-2 > 2$ where $\Upsilon$ is the helicity
modulus and this value has been confirmed both for relaxational and RSJ dynamics,
whereas $z=2\neq z_0$ was found for the same cases~\cite{beom}.
Consequently, the existence of two indices $z_0\neq z$ appears not to be restricted to relaxational dynamics. 

It is interesting to note that the $z_0$ value is more
 universal
 than the $z$: In 3D $z_0\approx 1.5$ is found for the LCG\cite{weber,wallin}
 and the $XY$ model (with FTBC) both for relaxational and RSJ dynamics.
 Likewise in  2D  $z_0(T)=2\pi\Upsilon(T)-2 $ is found for
 LCG\cite{weber1} (as well as for the 2D Coulomb Gas model with
 Langevin dynamics\cite{holmlund})
 and the $XY$ model (with FTBC) both in case of relaxational and RSJ dynamics.
On the other hand no such universality exists for $z$: In 3D we have the decreasing sequence
 $z\approx 2$, $1.5$, and $1$ when going from $XY$ relaxational dynamics,
 $XY$ RSJ dynamics, and LCG relaxational dynamics, respectively. The
 important point to note here is that the choice of dynamics imposed
 on the vortex lines leads to different dynamic characteristics even
 for the small $k$ and long time behavior. One may also observe that
 the $z$ value for the $XY$ model with relaxational dynamics is
 consistent with model A value $z\approx 2$ in the Hohenberg-Halperin classification
 scheme\cite{hohenberg} which to us suggests that the $z$ defined from
 $\tau\propto k^{-z}$ will in general be consistent with this
 classification scheme.

A crucial conclusion from the present investigation is that which of the two values is 
appropriate depends on the
experiment performed. For example a very common measurement for high $T_c$
superconductors is the $IV$ characteristics. In this experiment the
voltage across the sample is measured and consequently this
measurement relates to $z_0$. The scaling prediction for the
non-linear $IV$ characteristics gives $V\propto I^a$ where
$a=(z_0+1)/(d-1)$ in $d$ dimensions~\cite{ffh}.
A comparison with $a=(z+1)/(d-1)$ where $z$ is calculated according to the
Hohenberg-Halperin classification scheme is then likely to be incorrect
if the superconductor is described by 3D model A (because $z_0\neq z$
for this case), but would be correct
if it is described by 3D model F (because $z_0=z$ for this case~\cite{melwyn,nytt}).
Consequently the possibility of the existence of two critical
dynamic indices would have to be taken into account when analyzing
experiments. A criterion for when $z\neq z_0$ and $z=z_0$ remains to be found.

Support from the Swedish Natural Research Council through contracts
F 5102-659/2001 and F 650-19981449/2000 as well as support from the supercomputer center
HPC2N in Ume{\aa} are gratefully acknowledged.

\begin{figure}
\centering{\rotatebox{270}{\resizebox*{!}{7.0cm}{\includegraphics{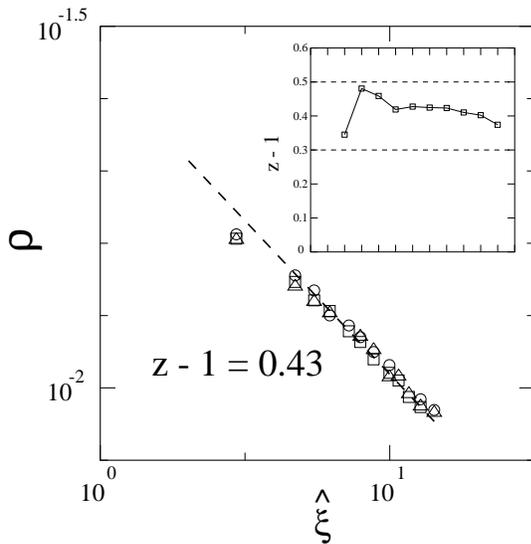}}}}
\caption{Determination of $z_0$ from $\rho\propto \xi^{1-z_0}$for the 3D
  LCG with PBC for lattice sizes
  $L$=10, 12,  and 16 (triangles, circles, and squares,
  respectively) and $\xi\propto \hat{\xi}\equiv (T-T_c)^{-\nu}$ (see text). As
  seen the data is independent of lattice size. The broken line is a
  least square fit to the linear part of the data and gives
  $z_0\approx 1.4(1)$ consistent with $z_0\approx 1.5$. The inset
  shows $z-1$ from the slope ($z-1=-$slope) for the $L=16$-data, obtained by
  least square fitting starting from the two data points for the
  largest $\hat{\xi}$ and then consecutively adding more data points
  so that the rightmost data point in the inset is based on all but
  one data point in the main figure. The dashed lines in the inset give our
  rough estimate of the error.}
\label{fig_1}
\end{figure}

\begin{figure}
\centering{\rotatebox{270}{\resizebox*{!}{7.0cm}{\includegraphics{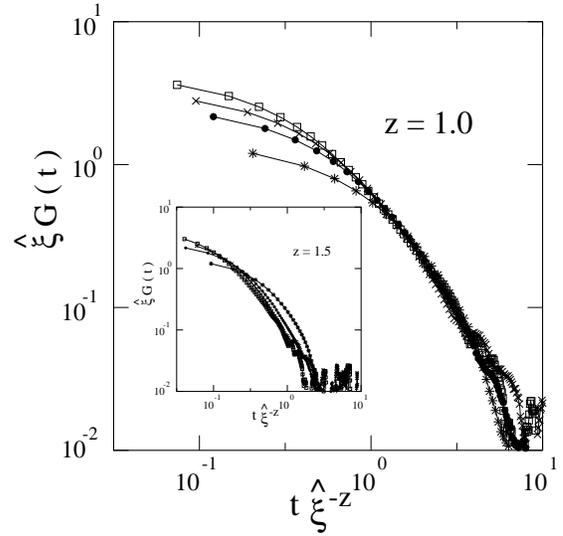}}}}
\caption{Determination of $z$ from the scaling $\xi G(0,t)\propto F(t\xi^{-z})$
  for the modified 3D LCG (corresponding to the 3D $XY$ model with
  PBC). The data are for lattice size $L=20$ and $T=$ 3.07, 3.10, 3.14, and 3.30
  (open squares, crosses, filled circles and asterisks, respectively). A good
  collapse is obtained for $z\approx 1$, whereas the inset shows that
  no collapse is obtained for $z=1.5$ ($\hat{\xi}$ is defined as in Fig.1). Consequently  
  $z\approx 1$, which is different from $z_0\approx 1.5$ obtained in Fig.1.}
\label{fig_2}
\end{figure}

\begin{figure}
\centering{\rotatebox{0}{\resizebox*{!}{6.0cm}{\includegraphics{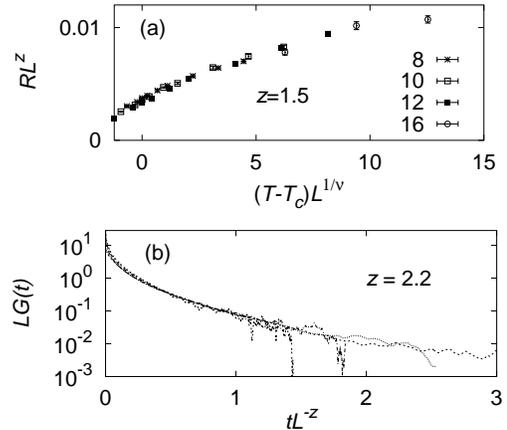}}}}
\caption{Demonstration that the 3D $XY$ model with FTBC contains two
  distinct indices $z$ and $z_0$ where $z\neq z_0$.
  (a) Scaling of resistance, $R\approx
  L^{-z_0}\tilde{F}(L(T-T_c)^{-\nu})$, close to and at $T_c$ gives
  $z_0\approx 1.5$. (b) Scaling $LG(k_{\rm min},t)=F(tk_{\rm min}^z ,
  tL^{-z_0})$ at $T_c$ gives $z\approx 2$ for large $t$  ($L=10, 12, 16$). Note that the
  scaling with the larger of $z$ and $z_0$ always dominates in the
  large $t$ limit.}
\label{fig_3}
\end{figure}

\end{multicols}

\newpage

\begin{figure}
\centering{\rotatebox{270}{\resizebox*{!}{7.0cm}{\includegraphics{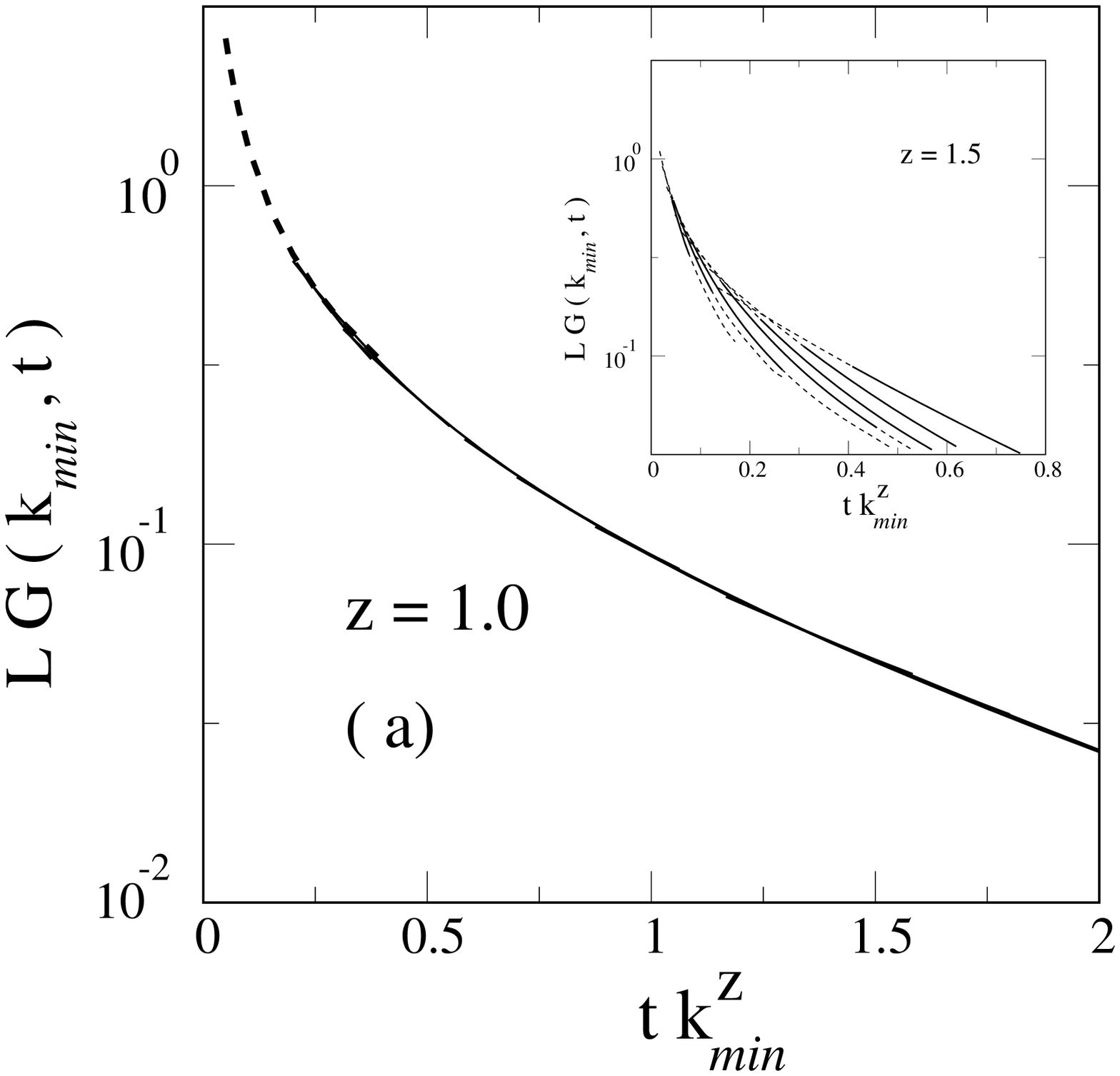}}}}
\centering{\rotatebox{270}{\resizebox*{!}{7.0cm}{\includegraphics{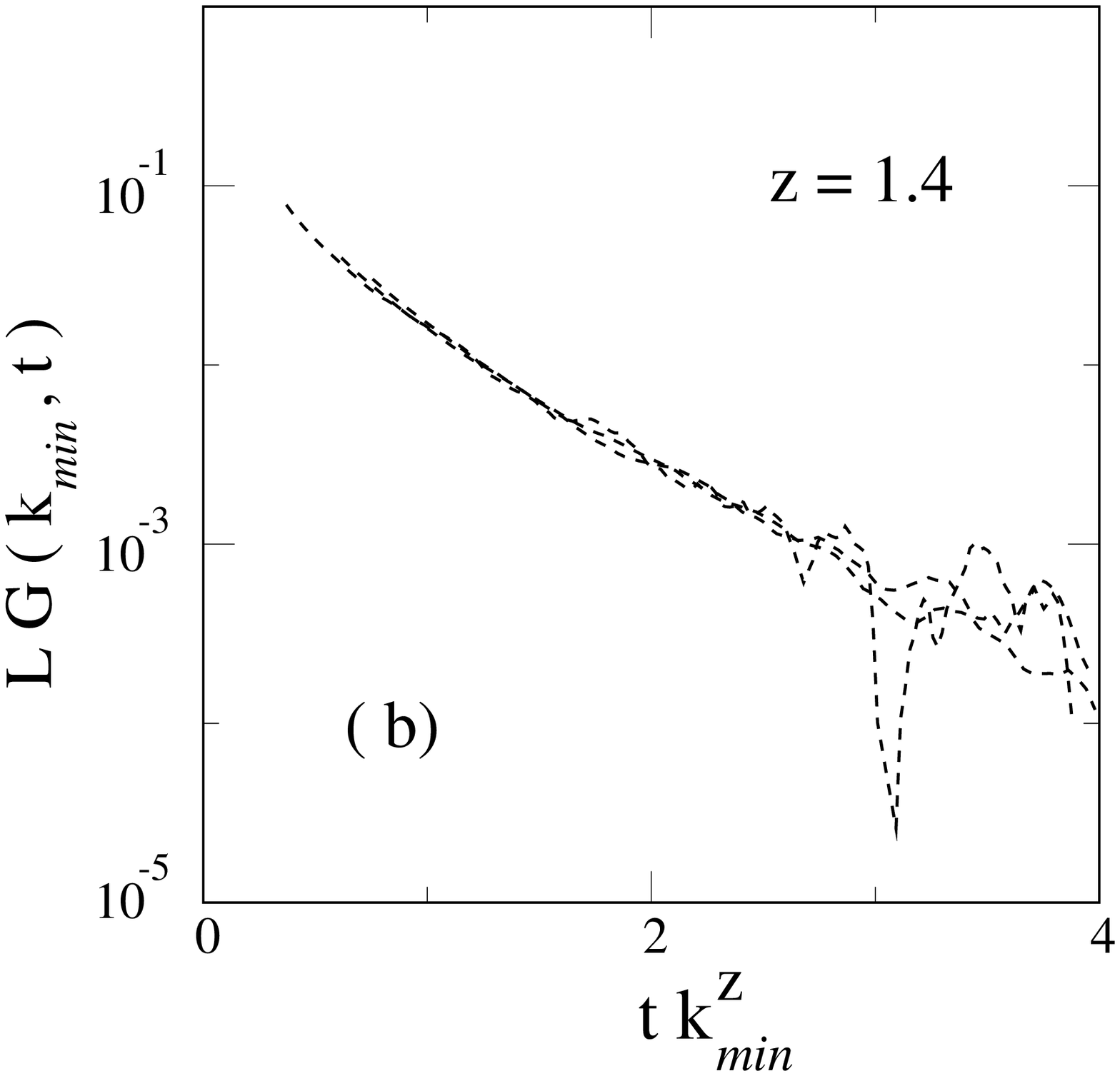}}}}
\caption{Demonstration that the 3D LCG with PBC at $T_c$ contains two distinct
  indices $z$ and $z_0$ within the single scaling function $LG(k_{\rm
    min},t)=F(tk_{\rm min}^z,tL^{-z_0})$.
  (a) The data are for $L$= 6, 8, 10, 12, 16, 20, and 24.
  The inset shows that $LG(k_{\rm min},t)$ does not scale with
  $tk_{\rm min}^{z=1.5}$ for the data shown (the value of $LG$ decreases for
    increasing $L$ for a fixed $tk_{\rm min}^{z=1.5}$). The full drawn
    part of the data for each size in the inset gives a middle section
    of the data corresponding to neither too large nor too small
    values of $t$. These full drawn middle sections of the data in the inset
    are the parts that collapse to a single scaling curve for
    $z\approx 1$, as
    demonstrated by the main part of the figure. The broken curve in
    the main part is the leading small $tk_{\rm min}^z$-dependence of
    the scaling function given by $C/tk_{\rm min}^z$, where $C$ is a
    constant. This leading term is consistent with the scaling curve
    obtained from the data ($C$ is used as an adjustable parameter).
    (b) Data for $L$=10, 12 and 16 obtained for larger $t$.
    These larger $t$  data collapse to a single curve with
    $z\approx 1.4$. Note that this is consistent with that the larger
    of $z$ and $z_0$ dominates in the large t limit.}
\label{fig_4}
\end{figure}


\begin{thebibliography}{99}
\bibitem{notez} The equality holds for a model where the critical
  exponent of the specific heat has a value $\alpha\leq 0$
  as for the 3D $XY$ model.
\bibitem{notescale} This limit simultaneously requires $t\gg \tau_0$ and
  $tL^{-z_0}\ll 1$. We believe that this requires
  converged data for larger $L$ than we could achieve.
\end{thebibliography}
\end{document}